\documentclass[twocolumn,showpacs,preprintnumbers,amsmath,amssymb]{revtex4}
\usepackage{graphicx}% Include figure files
\usepackage{dcolumn}% Align table columns on decimal point
\usepackage{bm}% bold math
\begin{document}
\title{Magnetization of a strongly interacting two-dimensional electron system in perpendicular magnetic fields}
\author{S. Anissimova, A. Venkatesan, A.~A. Shashkin$^*$, M.~R. Sakr$^\dag$, and S.~V. Kravchenko}
\affiliation{Physics Department, Northeastern University, Boston, Massachusetts 02115, U.S.A.}
\author{T.~M. Klapwijk}
\affiliation{Kavli Institute of Nanoscience, Delft University of Technology, 2628 CJ Delft, The Netherlands}
\begin{abstract}
We measure the thermodynamic magnetization of a low-disordered,
strongly correlated two-dimensional electron system in silicon in
perpendicular magnetic fields. A new, parameter-free method is used
to directly determine the spectrum characteristics (Land\'e
$g$-factor and the cyclotron mass) when the Fermi level lies outside
the spectral gaps and the inter-level interactions between
quasiparticles are avoided. Intra-level interactions are found to
strongly modify the magnetization, without affecting the determined
$g^*$ and $m^*$.
\end{abstract}
\maketitle

Magnetization is one of the least studied properties of
two-dimensional (2D) electron systems: signals associated with the
magnetization of 2D electrons are weak, and measuring them is a
challenging experiment. Few experimental observations of the
de~Haas–-van~Alphen effect in 2D electron systems were made using
SQUID magnetometers \cite{stormer83}, pick up coils lithographed
above the gate \cite{fang83}, or torque magnetometers
\cite{eisenstein85}. A novel method has recently been used by Prus
{\it et al}.\ \cite{prus03} and Shashkin {\it et al}.\
\cite{shashkin04a} to measure the spin magnetization of 2D electrons
in silicon metal-oxide-semiconductor field-effect transistors
(MOSFETs). This method entails modulating the magnetic field with an
auxiliary coil and measuring the imaginary (out-of-phase) component
of the AC current induced between the gate and the 2D electron
system, which is proportional to $\partial\mu/\partial B$ (where
$\mu$ is the chemical potential). Using the Maxwell relation,
$\partial\mu/\partial B=-\partial M/\partial n_s$, one can then
obtain the magnetization $M$ by integrating the induced current over
the electron density, $n_s$. Pauli spin susceptibility has been
observed to behave critically near the 2D metal-insulator transition,
in agreement with previous transport measurements
\cite{kravchenko04,shashkin04b}.

Here we apply a similar method to study the thermodynamic
magnetization of a low-disordered, strongly correlated 2D electron
system in silicon MOSFETs in perpendicular and tilted magnetic
fields. By measuring $\partial\mu/\partial B$ at non-integer filling
factors, we directly determine the spectrum characteristics without
any fitting procedures or parameters. As compared to previously used
measuring techniques, the remarkable advantage of the novel method is
that it probes the spectrum of the 2D electron system with the Fermi
level lying outside the spectral gaps so that the effects of
interactions between quasiparticles belonging to different energy
levels (inter-level interactions) are avoided. Although
intra-level interactions are found to strongly affect the
magnetization, the extracted Land\'e $g$-factor and the cyclotron
mass are insensitive to them. Therefore, measured spectrum
characteristics are likely to be identical with those of a continuous
spectrum. The so-obtained $g$-factor has been found to be weakly
enhanced and practically independent of the electron density down to
the lowest densities reached ($\approx1.5\times10^{11}$~cm$^{-2}$),
while the cyclotron mass becomes strongly enhanced at low $n_s$.

Measurements were made in an Oxford dilution refrigerator on clean
(100)-silicon samples with peak electron mobilities of 3~m$^2$/Vs at
0.1~K and oxide thickness of 149~nm. Magnetic field $B$ was modulated
with a small AC field $B_{\text{mod}}$ in the range of 0.005 --
0.03~T at a low frequency $f=0.05-0.45$~Hz to minimize possible
mechanical resonances and avoid overheating the sample. The latter
was verified by monitoring the temperature-dependent sample
resistance at $B=0$ and in the range of filling factors 1 to 6 with
and without modulation. Noticeably higher amplitudes $B_{\text{mod}}$
and/or frequencies $f$ caused overheating of the mixing chamber and
were avoided. The in-phase and out-of-phase components of the current
between the gate and the 2D electron system were measured with high
precision ($\sim10^{-16}$~A) using a current-voltage converter and a
lock-in amplifier. The imaginary (out-of-phase) current component is
equal to $\mbox{Im }i=(2\pi fCB_{\text{mod}}/e)\,d\mu/dB$, where $C$
is the capacitance of the sample. For measurements of the
capacitance, a similar circuit was used with a distinction that the
gate voltage was modulated and thus the imaginary current component
was proportional to the capacitance. The electron density was
determined from the capacitance oscillations.

\begin{figure}
\scalebox{0.48}{\includegraphics[clip]{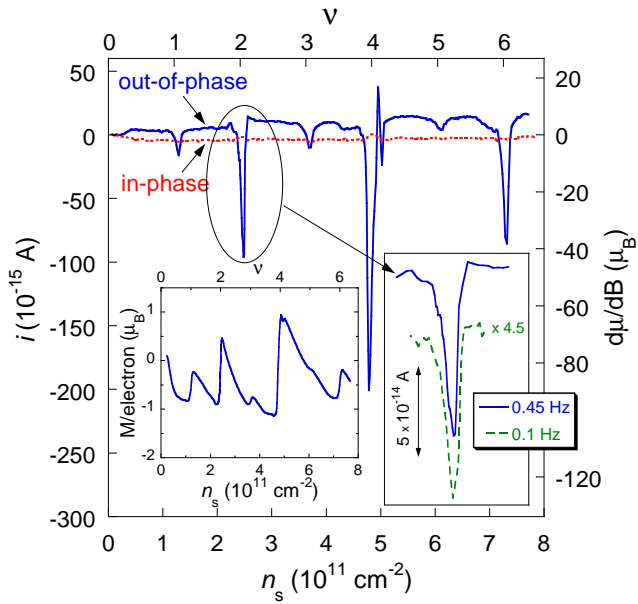}}
\caption{\label{fig1} Out-of-phase (solid line) and in-phase (dotted
line) current components as a function of the electron density in a
perpendicular magnetic field of 5~T and $T=0.8$~K.
$B_{\text{mod}}=0.022$~T and $f=0.45$~Hz. The value $d\mu/dB$ is
indicated in units of the Bohr magneton $\mu_B$. In the right-hand
inset, we demonstrate proportionality of $\mbox{Im }i$ to frequency:
the solid and dashed lines (vertically shifted for clarity)
correspond to $0.45$ and $0.1$~Hz, respectively; the $y$-component of
the latter is multiplied by 4.5. The left-hand inset illustrates
magnetization per electron.}
\end{figure}

Typical experimental traces of the gate current in a perpendicular
magnetic field of 5~T are displayed in Fig.~\ref{fig1}. Sharp dips in
the out-of-phase component, seen at integer filling factors
$\nu\equiv n_shc/eB_\perp$, reflect gaps in the density of states:
dips at odd filling factors correspond to the valley splitting, the
ones at $\nu=2$ and 6 are due to the spin splitting, and the dip at
$\nu=4$ is due to the cyclotron splitting. However, there are no
corresponding features in the in-phase current component, which
ensures that we reach the low-frequency limit and the measured
$\partial\mu/\partial B$ is not distorted by lateral transport
effects. This is further confirmed by the fact that the out-of-phase
current is proportional to the excitation frequency as displayed in
the right-hand inset to Fig.~\ref{fig1}. Magnetization per electron
can be extracted by integrating the measured out-of-phase signal with
respect to $n_s$, as shown in the left-hand inset to Fig.~\ref{fig1}
for illustration. The magnetization exhibits the expected sawtooth
oscillations, with sharp jumps at integer filling factors (note that
the height of the jumps yields values that are smaller than the level
splitting by the level width).

If the disorder and interactions are disregarded, in quantizing
magnetic fields (except at integer filling factors) the derivative
$\partial\mu/\partial B=-\partial M/\partial n_s$ is equal to
\begin{equation}
\frac{\partial\mu}{\partial B}=\mu_B\left[\left(\frac{1}{2}+N\right)\frac{2m_e}{m_b}\pm\frac{1}{2}g_0\right],
\end{equation}
where $\mu_B$ is the Bohr magneton, $N$ is the Landau level number,
$m_e$ and $m_b=0.19\,m_e$ are the free electron mass and band mass,
respectively, and $g_0=2$ is the $g$-factor in bulk silicon. Disorder
smears out the dependences which otherwise would consist of a series
of delta-functions. Interactions modify this picture in two ways:
(i)~by renormalizing the values of the cyclotron mass and $g$-factor
and (ii)~by providing a negative contribution of order
$-(e^2/\varepsilon l_B)\{\nu\}^{1/2}$ to the chemical potential
\cite{macdonald86,efros88} (here $\varepsilon$ is the dielectric
constant, $l_B$ is the magnetic length, and $\{\nu\}$ is the
deviation of the filling factor from the nearest integer). The latter
effect, which is caused by the intra-level interactions between
quasiparticles, leads to the so-called negative thermodynamic
compressibility near integer filling factors predicted by Efros
\cite{efros88} and experimentally observed in
Refs.~\cite{kravchenko89,eisenstein92}.

\begin{figure}
\scalebox{0.52}{\includegraphics[clip]{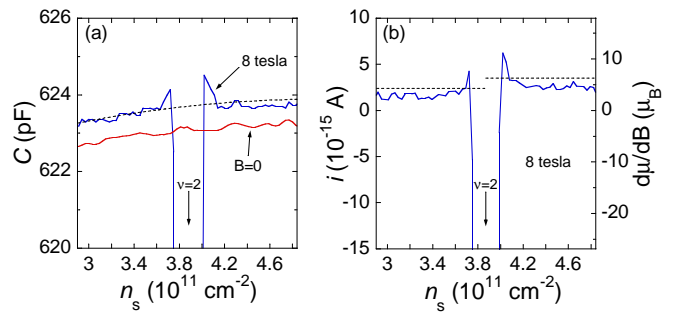}}
\caption{\label{fig2} (a)~Capacitance in $B=8$~tesla and in $B=0$ as
indicated. The (noise averaged) geometric capacitance is depicted by
a dashed line. (b)~$\mbox{Im }i\propto d\mu/dB$ in a perpendicular
magnetic field of 8~tesla. The maximum values possible in a
non-interacting system (see text) are depicted by dashed lines.}
\end{figure}

In Fig.~\ref{fig2}, we compare capacitance $C$ with
$\partial\mu/\partial B$, measured at the same magnetic field value
and plotted versus $n_s$ around the filling factor $\nu=2$. The
capacitance consists of two contributions:
$1/C=1/C_{\text{geo}}+1/Ae^2(dn_s/d\mu)$, where $C_{\text{geo}}$ is
the geometric capacitance \cite{rem} depicted by the dashed line in
Fig.~\ref{fig2}~(a), and $A$ is the sample area. (Note that the
geometric capacitance slightly increases with $n_s$ since the
thickness of the 2D electron layer --- and, therefore, the average
distance between the 2D layer and the gate --- decreases with the
gate voltage.) The second term is responsible for the dip centered at
$n_s=3.87\times10^{11}$~cm$^{-2}$, corresponding to $\nu=2$, and
sharp maxima on both sides of it. Note that at these maxima, the
capacitance exceeds $C_{\text{geo}}$, which corresponds to the
negative thermodynamic compressibility discussed above. Farther from
integer filling factors, the intra-level interaction corrections
become weak, being proportional to $\{\nu\}^{-1/2}$, and the measured
capacitance approaches $C_{\text{geo}}$ (as long as the broadening of
Landau levels is negligible, {\it i.e.},
$dn_s/d\mu\gg\left.dn_s/d\mu\right|_{B=0}$).

Similar maxima on both sides of $\nu=2$ are seen in the magnetization
data shown in Fig.~\ref{fig2}~(b). At the maxima, the derivative
$\partial\mu/\partial B$ exceeds maximum values possible in a
non-interacting 2D electron gas, which are determined by Eq.~(1) and
are depicted in the figure by dashed lines. The possibility that
$\partial\mu/\partial B$ might exceed its maximum non-interacting
values due to intra-level Coulomb interactions between quasiparticles
was predicted by MacDonald {\it et al}.\ \cite{macdonald86}; in fact,
this is how negative compressibility \cite{efros88} manifests itself
in magnetization measurements. Sharp spike just above $\nu=4$ and
maxima on both sides of $\nu=2$ in the dependence shown in
Fig.~\ref{fig1} are of the same nature.

\begin{figure*}
\scalebox{1.09}{\includegraphics[clip]{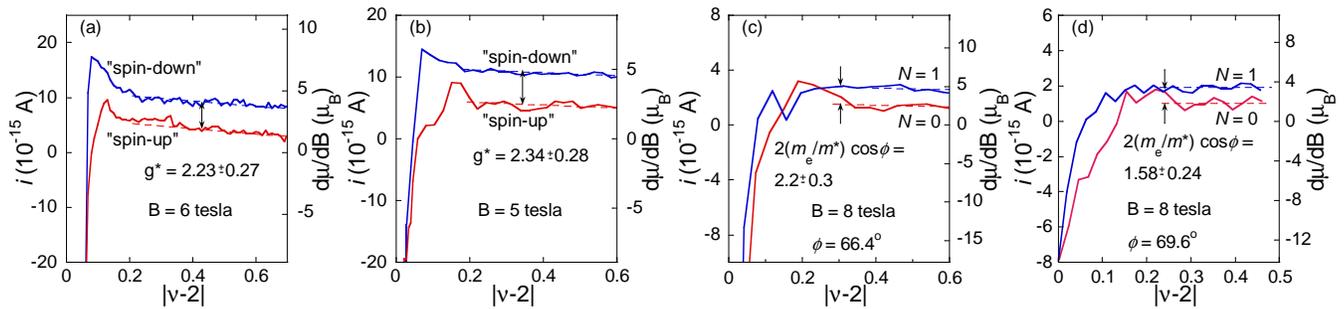}}
\caption{\label{fig3} Illustration of how the effective $g$-factor
(a, b) and the cyclotron mass (c, d) have been measured. The
imaginary current component is plotted as a function of the deviation
of the filling factor from $\nu=2$. In perpendicular magnetic fields,
the difference between $\partial\mu/\partial B$ for spin-down
($\downarrow$) and spin-up ($\uparrow$) electrons yields $g^*$ in
units of the Bohr magneton. In tilted magnetic fields, the difference
between $\partial\mu/\partial B$ for electrons with $N=1$ and $N=0$
is equal to $2\mu_B\,(m_e/m^*)\cos\,\phi$. The dashed lines show
noise-averaged values. $B_{\text{mod}}=0.022$~T (a, b) and 0.0055~T
(c, d).}
\end{figure*}

It is straightforward to obtain the effective $g$-factor from the
data for $\partial\mu/\partial B$. In accordance with Eq.~(1), it is
equal (in units of the Bohr magneton) to the difference between
$\partial\mu/\partial B$ for spin-down ($\downarrow$) and spin-up
($\uparrow$) electrons belonging to the same Landau level:
$\mu_B\,g^*= (\partial\mu/\partial
B)_\downarrow-(\partial\mu/\partial B)_\uparrow$. It is important
that this method of determining the $g$-factor does not require the
use of any fitting procedures or parameters. Figure~\ref{fig3}~(a, b)
shows measured $\partial\mu/\partial B$ as a function of the
deviation of the filling factor from 2 at two values of magnetic
field. Near $\nu=2$, there are sharp intra-level interaction-induced
structures discussed above; these regions have been excluded from the
analysis. However, farther from $\nu=2$, the dependences for $\nu<2$
and $\nu>2$ become parallel to each other. This ensures that the
so-determined $g^*$ is not affected by the valley splitting
\cite{hrapai03,valley} and intra-level interaction effects
\cite{macdonald86,efros88} discussed above. The latter contribute
equally to both spin-up and spin-down dependences and cancel each
other out. Disorder also contributes equally to $\partial\mu/\partial
B$ on both sides of $\nu=2$: we have found that at magnetic fields
down to approximately 3~T, there are wide regions of filling factors
where capacitance ({\it i.e.}, the density of states) is symmetric
around $\nu=2$ (see, {\it e.g.}, Fig.~\ref{fig2}~(a)); furthermore,
closeness of the capacitance to $C_{\text{geo}}$ attests that the
disorder-induced corrections are small. At lower magnetic fields,
however, the electron-hole symmetry around $\nu=2$ breaks down, which
sets the lower boundary for the range of magnetic fields (and,
consequently, electron densities). Note that temperature smears out
the dependences in a way similar to disorder: at higher temperatures,
the capacitance at half-integer filling factors decreases, which
leads to a worsening of the method accuracy.

\begin{figure}
\scalebox{0.52}{\includegraphics[clip]{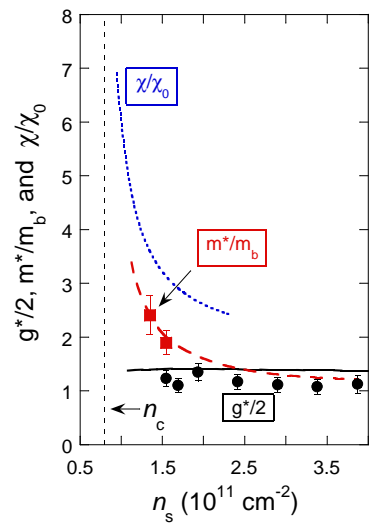}}
\caption{\label{fig4} The effective $g$-factor (circles) and the
cyclotron mass (squares) as a function of the electron density. The
solid and long-dashed lines represent, respectively, the $g$-factor
and effective mass, previously obtained from transport measurements
\cite{shashkin02}, and the dotted line is the Pauli spin
susceptibility obtained by magnetization measurements in parallel
magnetic fields \cite{shashkin04a}. The critical density $n_c$ for
the metal-insulator transition is indicated.}
\end{figure}

In Fig.~\ref{fig4} we plot the measured $g$-factor along with the one
previously obtained from transport measurements (solid line). One can
see that there is no systematic dependence of the $g$-factor on
$n_s$: it remains approximately constant and close to its value in
bulk silicon even at the lowest electron densities, which is in good
agreement with the transport \cite{shashkin02} and magnetocapacitance
\cite{hrapai03} results.

The same method can be used for determination of the cyclotron mass
in tilted magnetic fields strong enough to completely polarize the
electron spins \cite{remark}. If (and only if) the spin splitting
exceeds the cyclotron splitting, the gap at $\nu=2$ lies between
Landau levels 0$\uparrow$ and 1$\uparrow$, and the difference
$(\partial\mu/\partial B)_{N=1}-(\partial\mu/\partial B)_{N=0}$ is
equal to $2\mu_B\,(m_e/m^*)\cos\,\phi$, where $\phi$ is the tilt
angle. Once the electron spins are fully polarized at filling factors
above $\nu=2$, the tilt angle is automatically large enough for the
level crossing to have occurred. The region of explorable electron
densities is restricted from above by the condition that the
electrons must be fully spin-polarized, while with our current
set-up, the maximum magnetic field at which we can apply the
modulation is only 8~tesla capable of polarizing the electron spins
up to $n_s^*\approx2\times10^{11}$~cm$^{-2}$
\cite{shashkin04a,vitkalov00}. Figure~\ref{fig3}~(c, d) shows
$\partial\mu/\partial B$ as a function of $|\nu-2|$ under the
condition $n_s<n_s^*$ at two tilt angles \cite{angle}. The extracted
cyclotron mass at electron densities $1.55$ and
$1.35\times10^{11}$~cm$^{-2}$ is significantly enhanced. At densities
below $1.35\times10^{11}$~cm$^{-2}$, the symmetry of capacitance on
both sides of the $\nu=2$ gap breaks down, making the determination
of $m^*$ impossible. As a result, we were only able to obtain two
data points. Nevertheless, good agreement with the effective mass
previously obtained by transport measurements (Fig.~\ref{fig4})
demonstrates the applicability of the new method and adds credibility
to both transport and magnetization results.

We stress once again that the advantage of the new method we use here
is that it allows determination of the spectrum of the 2D electron
system under the condition that the Fermi level lies outside the
spectral gaps, and the inter-level interactions are avoided. Being
symmetric about $\nu=2$, the intra-level interactions are canceled
out in the data analysis and do not influence the extracted
$g$-factor and cyclotron mass. Therefore, the obtained values $g^*$
and $m^*$ are likely to be identical with those for a continuous
spectrum, and the comparison with previously found values of the
$g$-factor and the effective mass is valid.

To summarize, thermodynamic magnetization measurements in
perpendicular and tilted magnetic fields allow determination of the
spectrum characteristics of 2D electron systems and show that
enhancement of the $g$-factor is weak and practically independent of
the electron density, while the cyclotron mass becomes strongly
enhanced as the density is decreased. The obtained data agree well
with the $g$-factor and effective mass obtained by transport
measurements, as well as with the Pauli spin susceptibility obtained
by magnetization measurements in parallel magnetic fields, even
though the lowest electron densities reached in the experiment are
somewhat higher. Thus, we arrive at the conclusion that, unlike in
the Stoner scenario, it is indeed the effective mass that is
responsible for the dramatically enhanced spin susceptibility at low
electron densities.

We gratefully acknowledge discussions with V.~T. Dolgopolov, B.~I.
Halperin, and M.~P. Sarachik. This work was supported by the National
Science Foundation grant DMR-0403026, the ACS Petroleum Research Fund
grant 41867-AC10, the RFBR, RAS, and the Programme ``The State
Support of Leading Scientific Schools''.

{\it Note added in proof}.---After this work had been completed, we
learned that Punnoose and Finkelstein \cite{punnoose05} made a
renormalization group analysis for multi-valley 2D systems. Their
conclusion that the effective mass dramatically increases at the
metal-insulator transition while the $g$-factor remains nearly intact
is consistent with our experimental results.


\begin{thebibliography}{apssamp}
\bibitem{rem} The geometric capacitance has been determined according
to $1/C_{\text{geo}}=1/C|_{B=0}-1/Ae^2\left.(dn_s/d\mu)\right|_{B=0}$
(for more on this procedure, see Ref.~\cite{hrapai03}).
\bibitem{valley} The bare valley splitting is independent of the
magnetic field and does not contribute to $\partial\mu/\partial B$.
However, it may be enhanced by inter-level interactions
\cite{hrapai03}. The fact that $\partial\mu/\partial B$ for $\nu<2$
and $\nu>2$ are parallel to each other ensures that a possible
influence of the enhanced valley splitting is negligible in our
experiment.
\bibitem{remark} The effects of finite layer thickness, which lead to
an increase of the effective mass with parallel magnetic field, are
negligible in silicon MOSFETs \cite{kravchenko04,shashkin04b}.
\bibitem{angle} Using $n_s^*(8~\mbox{T})\approx2\times
10^{11}$~cm$^{-2}$, one can estimate that the coincidence of spin and cyclotron splittings for $\nu=2$ occurs at
$\phi^*=\cos^{-1}(n_s^*hc/2eB)\approx59^\circ$.
\end{thebibliography}
\end{document}